\documentclass[pra,twocolumn,notitlepage,superscriptaddress,longbibliography]{revtex4-1}
\usepackage{amsmath}
\usepackage{amssymb}

\usepackage[unicode=true,colorlinks=true,citecolor=blue,urlcolor=blue]{hyperref}

\usepackage{bm}
\usepackage{color}
\usepackage{epsfig}
\usepackage{multirow}

\usepackage[normalem]{ulem}
\renewcommand {\Im}{\mathop\mathrm{Im}\nolimits}

\renewcommand {\phi}{{\varphi}}
\newcommand {\rmi}{{\rm i}}

\newcommand {\e}{{\rm e}}
\newcommand {\eps}{\varepsilon}

\begin{document}
\title{%
%Topologically bound states, edge states and flat bands, induced by two-particle repulsion\\
Topologically bound states, non-Hermitian skin effect and flat bands, induced by two-particle interaction}

\author{Alexander N. Poddubny}
%\email{a.poddubny@fastmail.com}
\email{alexander.poddubnyy@weizmann.ac.il}

\affiliation{Rehovot 761001, Israel}
%\affiliation{Weizmann Institute of Science, Rehovot 7610001, Israel}

\begin{abstract}
We study theoretically quantum states of two repelling spinless particles in a   one-dimensional tight-binding model with simple periodic lattice and open boundary conditions. We demonstrate, that when the particles are not identical,  their interaction drives nontrivial correlated two-particle states, such as  bound states, edge states as well as interaction-induced flat bands. Specifically, the center-of-mass and relative motions of two particles become coupled in a topologically nontrivial way. By virtue of the non-Hermitian skin effect the localization of the center of mass enforces the localization of  the relative motion and  formation of the bound states. 
\end{abstract}
\date{\today}

%\cite{}

\maketitle

\section{Introduction} In the last decades topology-inspired ideas became a universal framework to characterize various natural   phenomena. First, topological excitations, solitons and vortices, were found in various systems with nonlinearity and interactions~\cite{manton2004topological}. Next, it was understood that even  non-interacting periodic systems, described by linear equations, can be assigned integer topological indices, and localized excitations arise at the boundaries where such indices exhibit an abrupt change~\cite{Hasan2010,Ozawa2019}. Such localized excitations, topological edge states, can form from  different species of particles, from electrons to photons to mechanical vibrations~\cite{Ma2019}. Even more recently   self-induced edge states with nontrivial topology assisted by nonlinearity became a subject of active studies~\cite{Poddubny2018Ring,alex2020quantum,Olekhno2020,Kirsch2021}.
Another aspect of the interplay  of topology and interactions that involves not only the edge states but also the bound states, has been  recently put forward in Refs.~\cite{Lee2021,Shen2022}.  It has been understood that topological nontrivial Hamiltonians may be realized also in the  systems of several interacting particles even without any physical boundaries. In this case one of the particles can provide a boundary where the other one can localize, thus forming a bound state. This localization mechanism has a  certain similarity with the one discussed in~Refs.~\cite{alex2020quantum,Zhong2020}, however,  it also involves the so-called non-Hermitian skin effect ~ \cite{Lee2016,Torres2018,Slager2020,Okuma2020}, that  means localization of the bulk eigenstates in the non-Hermitian system. As a result, the relative motion of the two particles with respect to each other becomes restricted.  One could  term such kind of states as {``topologically bound''}, in contrast to usual topological edge states.  Importantly, the bound states  considered in Refs.~\cite{Lee2021,Shen2022,Qin2022,Koh2022} involved also the non-reciprocity and/or non-Hermiticity of the system even in the absence of interaction.   

Here, we show that topologically bound states and a series of interaction-induced flat bands can be realized in an even simpler situation  of just two distinguishable interacting particles with different masses with open boundary conditions. At the single-particle level the model is reciprocal and Hermitian, while contact repulsion drives complex two-particle correlations and localizations.  All these two-particle correlations are just a consequence  of different masses of of the two particles. They do not require any special lattice engineering, as in a Su-Schrieffer-Heeger (SSH) model \cite{Gorlach2017,Olekhno2020,Salerno2020}, or even unusual long-ranged couplings   as  in the case of waveguide-coupled atom arrays~\cite{alex2020quantum,sheremet2021waveguide}. Hence, our results can be readily verified in a variety of quantum setups, such as lattices of trapped cold atoms, where bound states of two interacting identical particles have been observed~\cite{Winkler2006}. Importantly, the two-particle model with different masses has been considered before in detail
~\cite{Piil2008,Valiente2010,Valiente2019}. In Ref.~\cite{Valiente2019} even the formation of a single flat band has been predicted for the structure with an impurity. Here, however, instead of the impurity we consider open boundary conditions. Moreover, we obtain a series of flat bands with different energies instead of a single flat band.

The rest of the manuscript is organized as follows.
Section~\ref{sec:model} presents the model and outlines our main results. 
Section~\ref{sec:bound} contains the qualitative analytical argument for the formation of the topologically  bound state in this model. Next, in Sec.~\ref{sec:numerics}, we discuss the calculated eigenstates of the two-particle Schr\"odinger equation. Section~\ref{sec:Stark} details the origin of interaction-induced flat bands in our system and some details are reserved for Appendix~\ref{sec:Appendix}.

%%%%%%%%%%%%%%%%%%%%%%%%%%%%%%%%%%%%%%%%%%%%%%%%%%%%%
\begin
{figure}[t]
\centering\includegraphics[width=0.48\textwidth]{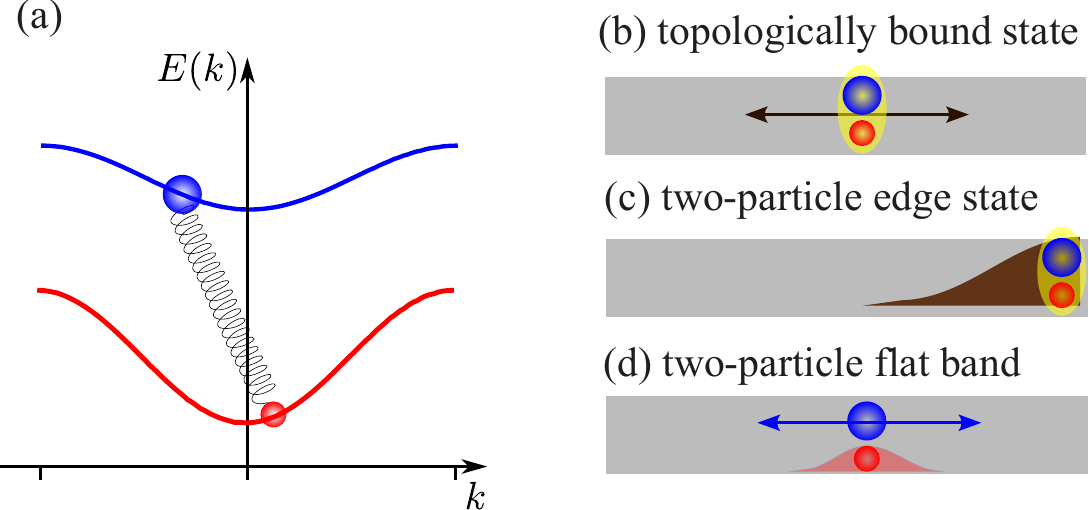}
\caption{(a) Schematics of the energy dispersion branches $E(k)$ for two distinguishable particles in a one-dimensional lattice. (b,c,d) Correlated states arising from a two-particle interaction: (b) {\it topologically bound} state of two-particles (c) two-particle edge state (d) two-particle flat-band state, where one of the particles is localized and the other one is delocalized. }
\label{fig:1}
\end{figure}
%%%%%%%%%%
\section{Model}\label{sec:model} We consider   a paradigmatic  one-dimensional (1D) tight-binding model of two different spinless particles 1 and 2, that exhibit a contact repulsion, as described by the Hamiltonian
\begin{multline}\label{eq:H}
H=\sum\limits_{\nu=1,2}\left[\sum\limits_{n=1}^N \eps_\nu b_n^{(\nu)\dag} b_n^{(\nu)}+
\sum\limits_{n=1}^{N-1} [t_\nu b^{(\nu)\dag}_nb^{(\nu)}_{n+1}+{\rm H.c.}]\right]\\+
U\sum\limits_{n=1}^{N} (b^{(1)\dag}_nb^{(1)}_n)(b^{(2)\dag}_nb^{(2)}_{n})\:.
\end{multline}
Each  of the particles is characterized by a site energy $\eps_{1,2}$ and a tunneling constant $t_{1,2}$, that yield  the dispersion laws $\eps_\nu(k)=\eps_{\nu}+2t_{\nu}\cos k$ in the absence of the interactions ($k$ is the quasimomentum). The corresponding dispersion curves are schematically shown in Fig.~\ref{fig:1}(a), their curvatures (particle masses) differ since $t_1\ne t_2$. Our goal is to examine the role of the interaction term, last line in Eq.~\eqref{eq:H}, on the two-particle correlations in the strong interaction regime when $|U|\gg |t_{1,2}|$. Namely, we demonstrate that for a finite number of sites $N$ this model has quite a rich structure of two-particle eigenstates  $|\Psi\rangle\equiv \sum_{nm} \psi_{nm}b_n^{(1)\dag} b_m^{(2)\dag}|0\rangle$, including  bound states  [Fig.~\ref{fig:1}(b)],  two-particle edge states [Fig.~\ref{fig:1}(c)] and also unusual two-particle correlated states where  one of the particles is localized and the second one is not [Fig.~\ref{fig:1}(d)]. We show that such two-particle states are degenerate with respect to the heavier particle position and form a series of flat bands~  \cite{Rhim2021}.

%%%%%%%%%%%%%%%%%%%%%%%%%%%%%%%%%%%
\section{Topologically bound states}\label{sec:bound}
%%%%%%%%%%%%%%%%%%%%%%%%%%%%%%%%%%%
\begin{figure}[t]
\centering\includegraphics[width=0.48\textwidth]{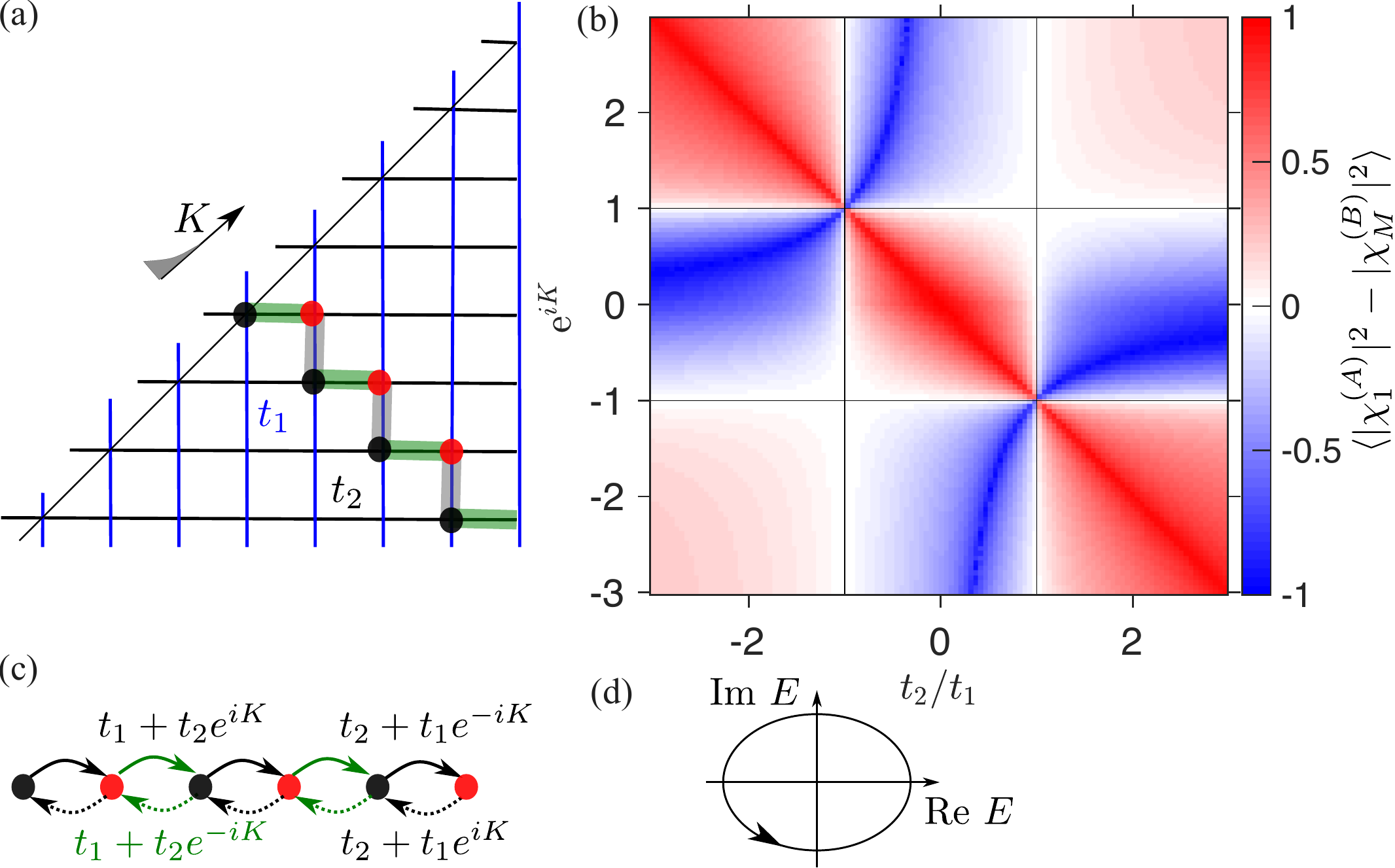}
\caption{(a) Scheme of the interaction of two particles with the center-of-mass wave vector $K$. (b) Interaction-induced  Su-Schrieffer-Heeger model with the couplings indicated above the arrows. (c) Localization parameter $\langle |\psi_1|^2-|\psi_N|^2\rangle$ depending on the ratio of the tunneling constants $t_2/t_1$ and on the center-of-mass wave vector $K$. (d) Winding of the complex energies Eq.~\eqref{eq:Ekappa} in the bulk around the coordinate origin.}
\label{fig:SSH}
\end{figure}
%%%%%%%%%%%%%%%%%%%%%%%%%%%%%%%%%%%%%%%%%%%%%%%%%%%%%
We start the analysis of the model by writing the two-particle Schr\"odinger equation for the wavefunction
%\begin{equation}
$\psi_{nm}=\chi_{n-m}\exp[\rmi K(n+m)]$,
%\end{equation}%
where $K$ is the center-of-mass wave vector and the amplitude $\chi$ characterizes the relative motion. For simplicity we always consider a situation where  the order of particles is fixed, that is either $n>m$ or $n<m$. The interacting two-particle model is then formally equivalent to the single-particle model on the right-angular discrete billiard in 2D, as shown in Fig.~\ref{fig:SSH}(a). Such discrete billiard was recently analyzed in detail in Ref.~\cite{Vidmar2022}, but neither bound nor localized states were considered there.  Vertical and horizontal motion  in the 2D lattice correspond to the motion of two particles.  Importantly, in the relative reference frame such  model  corresponds to a zigzag chain with  two sites per unit cell, that we will label A and B [thick line in Fig.~\ref{fig:SSH}(a)].
The Schr\"odinger equation for the amplitude  $\chi_n$ assumes the form
\begin{align}\label{eq:SSH}
\eps\chi_n^{(A)}&=(t_2+t_1/z)\chi_n^{(B)}+(t_1+t_2/z)\chi_{n-1}^{(B)}\:,\\
\eps\chi_n^{(B)}&=(t_2+t_1z)\chi_n^{(A)}+(t_1+t_2z)\chi_{n+1}^{(A)}\:,\nonumber
\end{align}
where $z=\e^{\rmi K}$ and $n=1,2\ldots$.
As soon as the center of mass motion is evanescent, $\Im K\ne 0$,   the system Eq.~\eqref{eq:SSH} realizes a non-Hermitian Su-Schrieffer-Heeger model 
\cite{Okuma2022,Bergholtz2021}, illustrated  in Fig.~\ref{fig:SSH}(b). Due to the non-Hermiticity such  model features the  non-Hermitian skin effect~ \cite{Lee2016,Torres2018,Slager2020,Okuma2020}:  all its eigenstates can become   localized at the edge. This  feature is specific for a  non-Hermitian system and  it is  related to the notrivial winding number of the eigenvalues of Eq.~\eqref{eq:SSH} in the bulk for a periodic solution of the form $\chi_n\propto \e^{\rmi \kappa n} $, where  $\kappa$ is the eigenvector~\cite{Leykam2017}. The complex eigenenergy is given by
\begin{multline}\label{eq:Ekappa}
E(\kappa)=\pm \sqrt{t_2+t_1/z+(t_1+t_2/z)\e^{-\rmi \kappa}}\\\times\sqrt{
t_2+t_1z+(t_1+t_2z)\e^{\rmi \kappa}}\:.
\end{multline}
It can be directly checked that for $z\ne 1$, $t_2\ne t_1$ Eq.~\eqref{eq:Ekappa} winds once around the point $E=0$ as $\kappa$ changes from $-\pi$ to $\pi$, as shown in Fig.~\ref{fig:SSH}(d).  In order to see how the eigenstates become localized one can e.g. substitute $\chi_n^{(B)}=0$ into the second of Eqs.~\eqref{eq:SSH} and find 
\begin{equation}\label{eq:skin}
\chi_n^{(A)}\propto \left(\frac{t_1z+t_2}{t_2z+t_1}\right)^n\:.
\end{equation}
Thus, as soon as $|t_1|\ne |t_2|$ and $|z|\ne 1$, the states become localized which turns out to be a generic topological feature. In order to better illustrate this
we have plotted numerically the localization parameter $|\chi^{(A)}_1|^2-|\chi^{(B)}_N|^2$, averaged over all the eigenstates, for a finite number of unit cells $n=1\ldots N=20$ depending on the ratio  $t_1/t_2$ and on the center-of-mass parameter $z$. The calculation demonstrates formation of localized states either at the left edges (red shading) or at  the right edges (blue shading), in agreement with Eq.~\eqref{eq:SSH}.

Our analysis thus predicts that for two different particle masses their center-of-mass motion, described by $K$, and their relative motion, described by $\chi$, are not independent but rather coupled  in a topologically nontrivial way. By virtue of the non-Hermitian effect the localization of the center of mass enforces constraints on  the relative motion, i.e. formation of the bound states.
Next, we will show by a rigorous numerical calculation of the two-particle eigenstates that center of mass can be localized indeed in a finite 1D array and that bound states form.

%%%%%%%%%%%%%%%%%%%%%%%%%%%%%%%%%%%
\section{Numerical results}\label{sec:numerics}
%%%%%%%%%%%%%%%%%%%%%%%%%%%%%%%%%%%
%%%%%%%%%%%%%%%%%%%%%%%%%%%%%%%%%%%%%%%%%%%%%%%%
\begin{figure}[t]
\centering\includegraphics[width=0.48\textwidth]{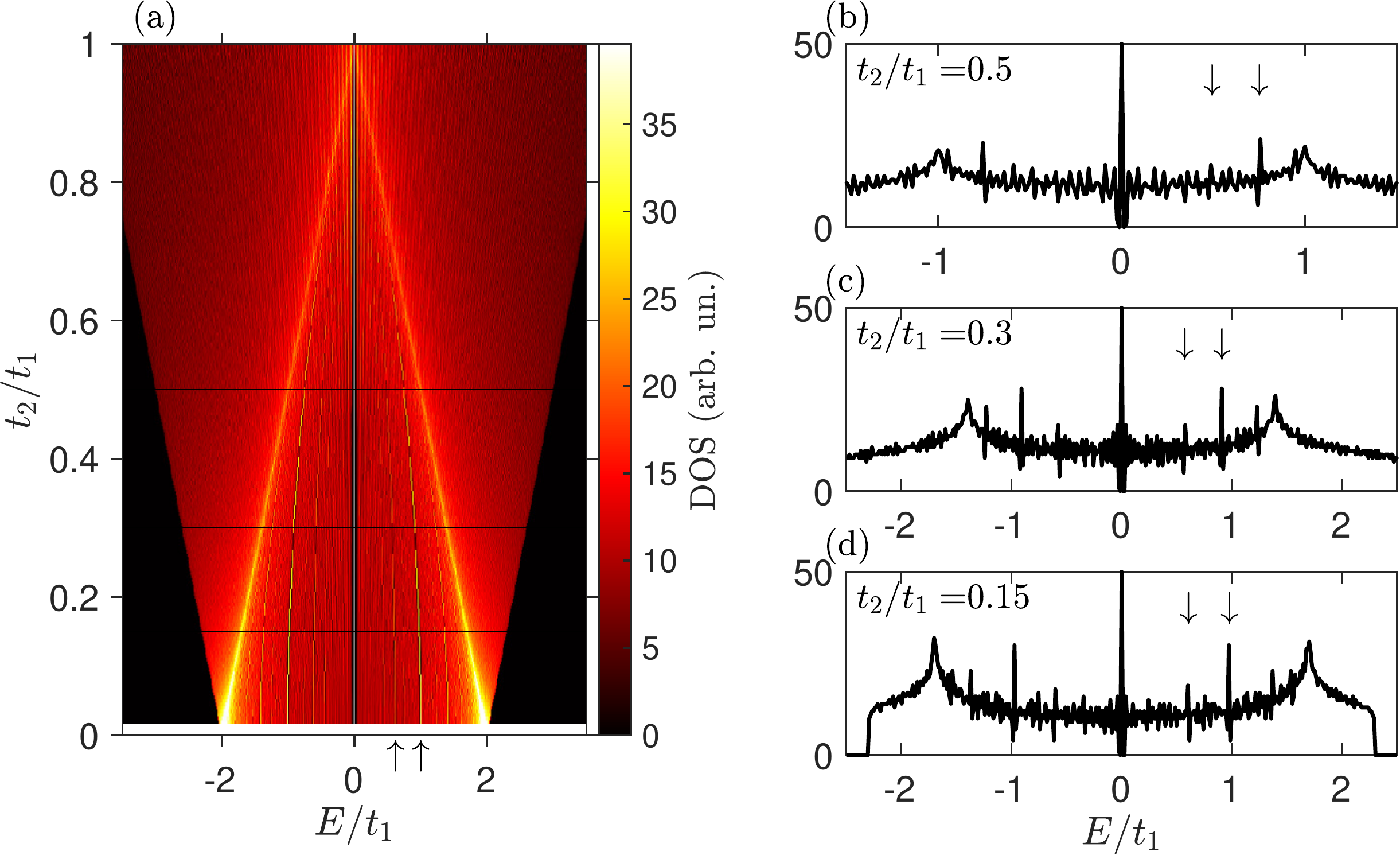}
\caption{(a) Density of states depending on the value of the tunneling constants $t_2/t_1$ for $N=101$. (b,c,d)
Cross-sections of (a) for three different values of $t_2/t_1$, indicated on graphs and also shown  by horizontal lines in (a).}
\label{fig:DOS}
\end{figure}
%%%%%%%%%%%%%%%%%%%%%%%%%%%%%%%%%%%%%%%%%%%%%%%%
%%%%%
\begin{figure}[t]
\centering\includegraphics[width=0.48\textwidth]{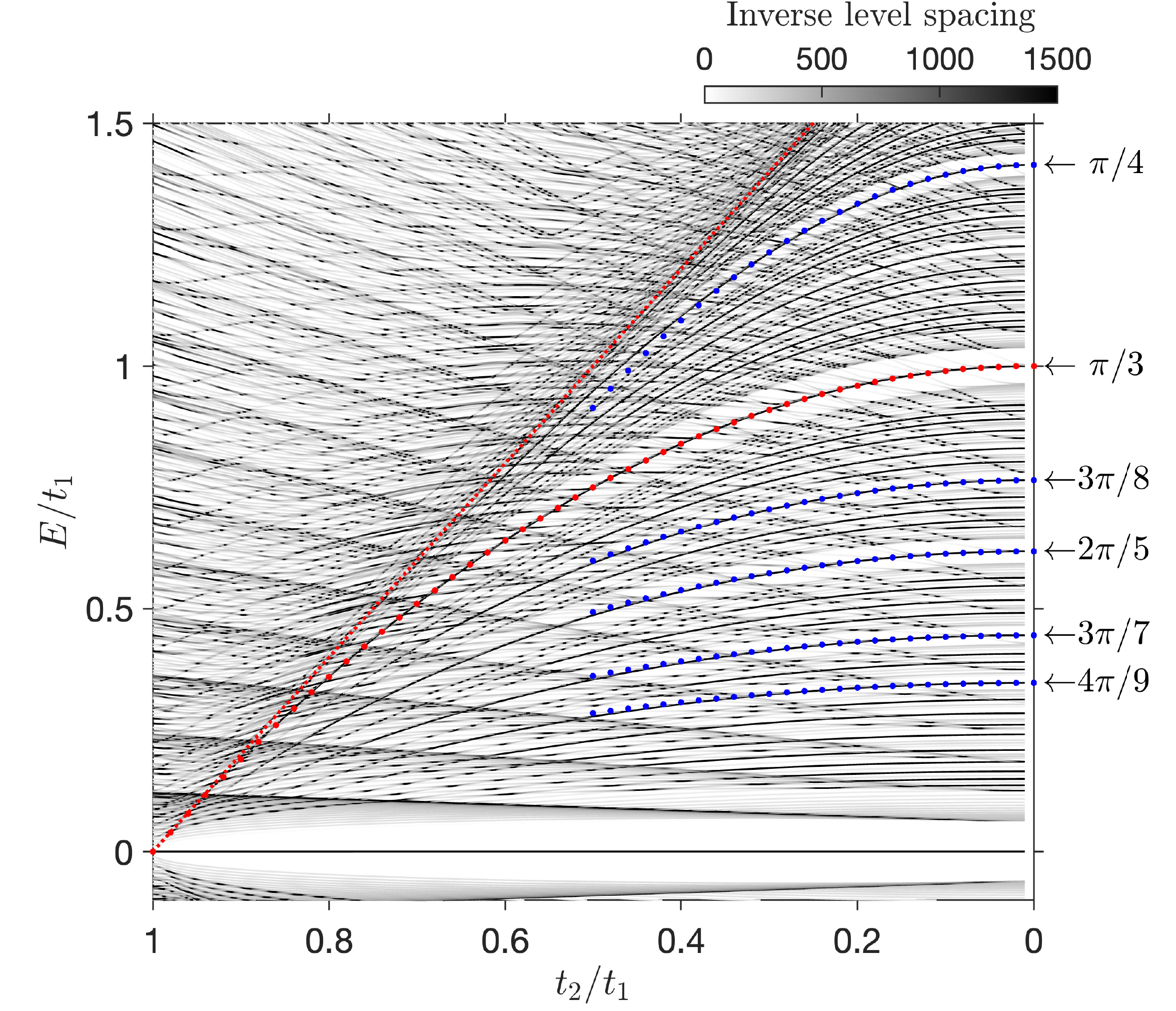}
\caption{Energy of states  depending on the value of the tunneling constants $t_2/t_1$ calculated for $N=101$ and $t_1=1$. The grayscale intensity corresponds to the inverse spacing between neigboring levels. Dotted line shows the dependence $E=2(t_1-t_2)$. Horizontal arrows  indicate  the energies  $E=2t_1\cos k$, with the values of $k$ given by each arrow. Dotted parabolas show the energies of the corresponding states calculated up to the second order in $t_2$.}
\label{fig:levels}
\end{figure}
%%%%%
\begin{figure*}[t]
\centering\includegraphics[width=\textwidth]{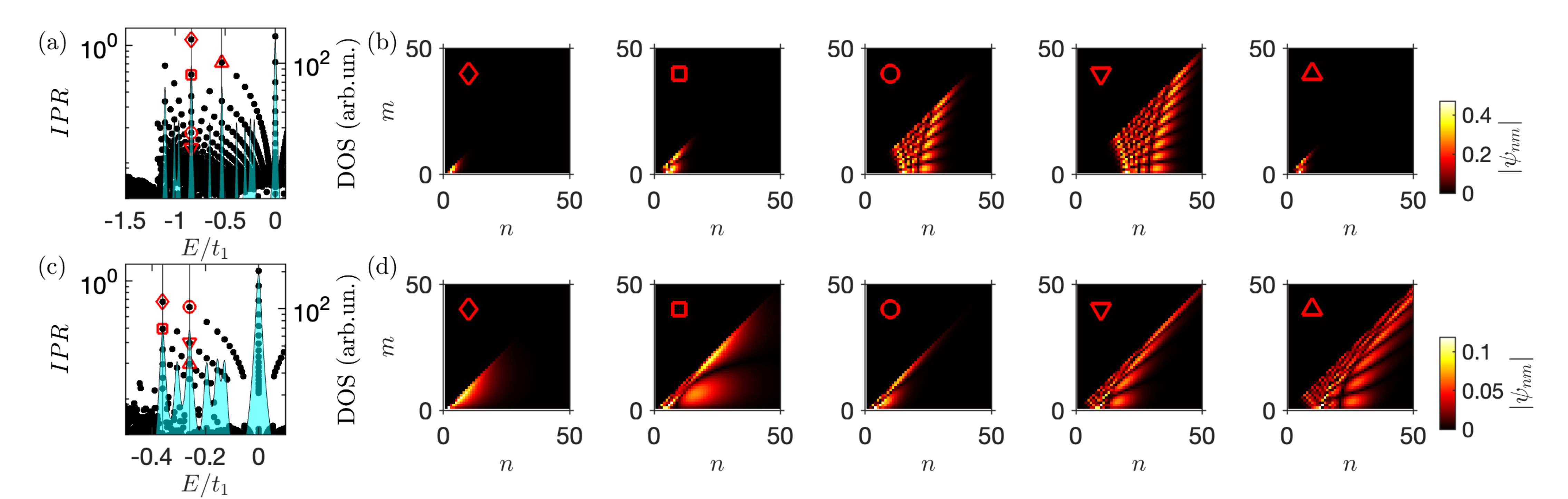}
\caption{(a,c) Energies of states vs inverse participation ratio (IPR) calculated for $t_2/t_1=0.4$ (a) and 
$t_2/t_1=0.8$ (c). Open symbols indicate specific eigenstates, with the corresponding wavefunctions shown in (b) and (d). Right $y$-axis shows by the blue shaded curves the density of states for all the states with the IPR larger than 0.3.}
\label{fig:eig}
\end{figure*}
%%%%%%%%%%%%%%%%%%%%%%%%%%%%%%%%%%
We present the two-particle state as
$
\Psi=\sum_{n=1}^N\sum_{m=n+1}^N \psi_{nm} b_n^{(1)\dag}b_m^{(2)\dag}|0\rangle\:,
$
and solve numerically  the Schr\"odinger equation $H\Psi=E \psi$ with the Hamiltonian Eq.~\eqref{eq:H} for the two-particle amplitudes $\psi_{nm}$ in the limit $U\to \infty$. We use $\eps_1+\eps_2$ as a reference point for the energy, which is equivalent to setting  $\eps_{1,2}=0$. 
Figure~\ref{fig:DOS} presents the density of states (DOS) numerically calculated for varying ratio $t_2/t_1$. 
This ratio  of the two tunnelling constants $t_2/t_1$  turns out to be the key parameter of the model.
Panel (a) shows DOS as a color map, and panels (b--d) show the plots of DOS vs. the energy $E$ for three characteristic ratios of $t_2/t_1$. Importantly, DOS is a strongly nonmotonous function of energy. First, it features  van Hove   singularities at the energies $E=\pm 2(t_1-t_2)$, corresponding to the extrema of single-particle dispersions $2t_{1,2}\cos k$. These singularities manifest themselves as sharp maxima that are best seen for $t_2=0$, when they are located at $E=\pm 2t_1$, see the bottom of  Fig.~\ref{fig:DOS}(a) and Fig.~\ref{fig:DOS}(d). However, there also exist additional sharp peaks in the DOS, not associated with the van Hove singularities. These extra sharp peaks are indicated by two vertical arrows in  each of the panels of Fig.~\ref{fig:DOS}. They correspond to the almost-degenerate states and arise from the interplay of the discreteness of the system and the interactions.  The presence of such peaks can be also seen in  Fig.~\ref{fig:levels} where we show the same  energy levels separately. In order to highlight the presence of degenerate states, corresponding to the interaction-induced flat bands, the intensity of the grayscale shading represents the inverse level spacing. In such way the degenerate states become brighter and stand out. Several sets of such states are highlighted by the dotted parabolas, that correspond to the DOS peaks discussed above. The two arrows in Fug.~\ref{fig:DOS} correspond to the parabolas labelled as $\pi/3$ and $2\pi/5$ in Fig.~\ref{fig:levels}. The origin of such notation will be discussed in Sec.~\ref{sec:Stark}. 
The eigenstates $\psi_{nm}$ at  these DOS peaks feature unusual two-particle correlations.

Figure~\ref{fig:eig} examines in   detail the spatial profile of the eigenstates for  $t_2/t_1=0.4$ (a,b) and $t_2/t_1=0.8$ (c,d).  Left panels (a,c) present the inverse participation ratio (IPR), that is defined as 
%\begin{equation}
$IPR=\sum_{nm}|\psi_{nm}|^4/(\sum_{nm}|\psi_{nm}|^2)^2$
%\end{equation}
depending on the eigenstate energy. The larger is IPR the stronger is the eigenstate localization. This calculation shows that a lot of the eigenstates are almost degenerate:  the points  in Figs.~\ref{fig:eig}(a,c) lie on the same vertical lines. The same is evidenced by the  density of states peaks, shown in Figs.~\ref{fig:eig}(a,c) by the   curves with blue shading, corresponding to the right abscissa axis. 
 The two vertical lines in Fig.~\ref{fig:eig} indicate  the same peaks in the density of states that are denoted by the arrows in Fig.~\ref{fig:DOS}. 
We also note the presence of a large central peak at $E=0$. This peak is associated with the chiral symmetry of the problem and corresponds to the states with a checkerboard profile,  localized on only part of the sites of the square lattice~\cite{Vidmar2022}. In this work, however, we are  interested in the states with $E\ne 0$.

Right panels Fig.~\ref{fig:eig}(b,d) show the spatial profiles $|\psi_{nm}|^2$ for several characteristic two-particle states, indicated by symbols in Fig.~\ref{fig:eig}(a,c). The abscissa and ordinate on these plots correspond to the coordinates of the two particles.  Crucially, the spatial distribution of the two-particle correlations is highly inhomogenous. For example, first two states in Fig.~\ref{fig:eig}(b) (diamond and square symbols), correspond to both particles localized at the edge of the lattice. This is a two-particle edge state,  akin to Fig.~\ref{fig:1}(c).
Next two states (circle and down-pointing triangle)  realize the situation when one particle is relatively localized and the other one is spread over the whole lattice. This is the state in Fig.~\ref{fig:1}(d). There exist many such states that differ only by a position of the localized particle. Since they are degenerate, they can be seen as an interaction-induced flat band. The last state in Fig.~\ref{fig:eig}(b) (upward-pointing triangle) is also a two-particle edge state, but with a different energy.

As the ratio of the tunnelling constants increases to $t_2/t_1=0.8$, the relative motion of the particles becomes also constrained. Namely, first three states in  Fig.~\ref{fig:eig}(c) (diamond, square and circle symbols) can be interpreted as the two-particle bound states, localized at the edge. The last two states (triangles) are  of somewhat intermediate character. They resemble both the above-mentioned state where only one of the particles is localized and also the two-particle bound state.
%%%%%%%%%%%%%%%%%%%%%%%%%%%%%%%%%%%%%%%%%%%%%%%%%%%%%
\begin{figure}[t]
\centering\includegraphics[width=0.45\textwidth]{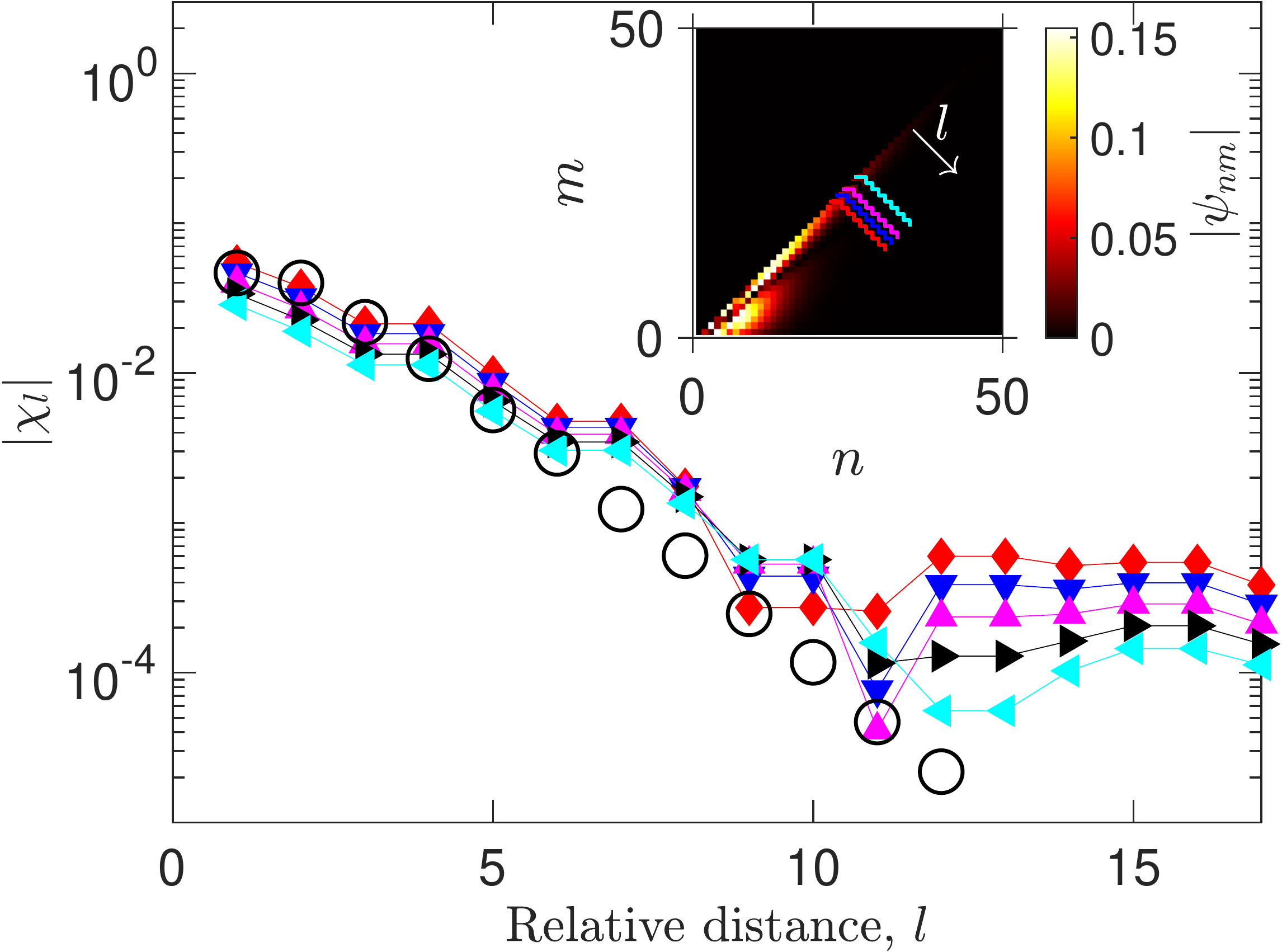}
\caption{Two-particle wavefunction of the ``topologically bound'' state depending on  the distance between the two-particles $l$.
Filled triangles with different orientation correspond to the different center-of-mass coordinates $n=22\ldots 26$. The points have been obtained by cutting the  total wave function $|\psi_{mn}|$ in the direction of the relative motion along the  colored zigzag lines, shown in the inset.
Open circles correspond to the eigenstate of the system Eq.~\eqref{eq:SSH} calculated for $z\equiv\e^{\rmi K}=-0.85$ for the most localized state with $E\approx -0.259 t_1$, that is indicated  by a circle in Fig.~\ref{fig:eig}(c,d).
Other calculation parameters are  $t_2/t_1=0.8$ and $N=101$.
}
\label{fig:skin}
\end{figure}
%%%%%%%%%%%%%%%%%%%%%%%%%%%%%%%%%%%%%%%%%%%%%%%%%%%%%

Thus, our rigorous numerical calculations confirm the formation of two-particle bound states, localized at the structure edge, in agreement with the non-Hermitian SSH model in Fig.~\ref{fig:SSH}(b). In order to further support the connection of the formation of a bound state with the non-Hermitian skin effect we  analyze in Fig.~\ref{fig:skin}  in more detail the eigenstate in Fig.~\ref{fig:eig}(d), indicated by a circle.  In the main part of Fig.~\ref{fig:skin} we present the cuts of the  distribution $|\psi_{nm}|^2$ along  the direction corresponding to the relative motion of the two particles. The specific points used for the cuts are indicated by the zigzag lines in the inset, that also shows the same spatial distribution $|\psi_{nm}|^2$. All of the cuts decay with the distance which reflects the confinement of the two particle to each other.  The difference between the cuts corresponds to the localization of the center of mass at the edge of the structure. Fitting this decay we were able to extract the center-of-mass localization parameter $z\equiv \exp(\rmi K)\approx -0.85$ for this eigenstate. Next, we have used this parameter in the effective non-Hermitian SSH model Eq.~\eqref{eq:SSH}. The resulting distribution of the eigenstate  of the effective model with the closest energy is shown in Fig.~\ref{fig:skin} by open circles. This state decays in space and the scale of the decay  satisfactory reproduces  the results of full numerical calculation (filled symbols). This agreement supports our interpretation of the formation of the bound state as a result of a non-Hermitian skin effect for the  relative motion.

The fact that such bound states arise only for relatively high values  of $t_2/t_1$, close to unity, also well agrees with the non-Hermitian SSH model. It is clearly seen in Fig.~\ref{fig:SSH}(c) that the darker color, that means stronger localization, corresponds to the regions with $t_2/t_1$ close to unity.

The analysis above  leaves two more open questions. First, what is the specific origin of the center-of-mass localization at the edge? Second, how and why form the flat bands, i.e. almost degenerate eigenstates? These questions turn out to be related and will be addressed in the next section.

\section{Interaction-induced Wannier-Stark ladder}\label{sec:Stark}
The formation of degenerate and localized states can be most easily understood in the limit when $t_2\ll t_1$, that is when one of the two particles is much heavier than the other one. In this case the motion of the heavier  particle can be considered as a perturbation. This means that  the triangular lattice in Fig.~\ref{fig:SSH} separates into vertical columns of varying height (blue color). One can first find the eigenstates within each column and then describe the coupling between the columns $\propto t_2$. The column height variation imposes an additional potential,   depends on the column height. Such model with coupled columns  is in fact very similar to the behavior of the particle on a 1D lattice in a constant electric field, that is described by a Wannier-Stark ladder~\cite{Fukuyama1973,Mendez1993,Gluck2002}. 

In order to derive such a model we start by writing  the Schr\"odinger equation in a tight-binding model for every column, that is 
\begin{equation}
t_1(\psi_{m-1,n}+\psi_{m+1,n})=\eps \psi_{m,n}
\end{equation}
with the open boundary conditions, $m=1\ldots n-1$.
Its eigenstates are the usual standing waves
\begin{equation}\label{eq:standing}
\psi_{mn}^{(\nu)}\approx \sqrt\frac{2}{n-1}\sin k_n^{(\nu)}m, \text{ where } k_x^{(j)}=\frac{\pi j}{n}\:, 
\end{equation}
$m,\nu=1,2,\ldots n-1$,with the energies
\begin{equation}
\eps_{n}^{(\nu)}=2t_1\cos k_n^{(\nu)}.
\end{equation}
Importantly, for each $n=3,6,\ldots$, that is divisible by $3$, there exists  an eigenstate with 
an integer $\nu=n/3$ so that $\cos k_n^{(\nu)}=1/2$ and $\eps_n^{(\nu)}=t_1$.
%\begin{equation}
%x=3m-1+dx, k_x^{(j)}=\frac{\pi j}{3j+dx}\approx\frac{\pi}{3}-\frac{\pi dx}{9j},\quad 
%E_{xj}\approx 1+\frac{\sqrt{3}\pi dx}{9j}
%\end{equation}
We now take into account the coupling between the states Eq.~\eqref{eq:standing}, i.e. the interaction between the ``columns'' that is proportional to the tunnelling constant $t_2$:
\begin{multline}\label{eq:coupling}
\langle \nu, n|\nu', n+1\rangle\equiv  t_2\sum\limits_{m=1}^{n-1}\psi_{m,n}^{(\nu)}\psi_{m,n+1}^{(\nu')} \\=-t_2\frac1{\sqrt{n(n-1)}}\frac{\sin k_n^{(\nu)}\sin k_{n+1}^{(\nu')}}{\cos k_n^{(\nu)}-\cos k_{n+1}^{(\nu')}}\:.
\end{multline}
We can now formally write  a coupled-columns model
\begin{multline}
\eps_n^{(\nu)} \psi_{n}^{(\nu)}+\langle \nu, n|\nu', n+1\rangle \psi_{n+1}^{(\nu)}+
\langle \nu, n|\nu', n-1\rangle \psi_{n-1}^{(\nu)}\\=\eps_n^{(\nu)} \psi_{n}^{(\nu)}\:.
\end{multline}
This is equivalent to rewriting the original two-particle Schr\"odinger equation into the standing-wave basis only for one of the particles.

We  now  assume that $\nu=n_0/3$ with $n_0\gg 1$ and take into account only one standing wave $\nu$. Next, we expand the matrix element Eq.~\eqref{eq:coupling} in the limit where $k^{(\nu)}\approx \pi/3$. This results in the following equation 
\begin{equation}\label{eq:stark}
[F(n-n_0)+\alpha (n-n_0)^2]\psi_n+\tau (\psi_{n+1}+\psi_{n-1})=\varepsilon\psi_n\:, 
\end{equation}
with $\varepsilon=E/t_1-1$, $F=\pi/\sqrt{3}n_0$ being the dimensionless electric field,
$\alpha=-\pi(6\sqrt{3} + \pi)/18n_0^2$ and
$\tau=3\sqrt{3} t_2/(2\pi t_1)$. The index $\nu$ is dropped for simplicity.  Contrary to the classical Wannier-Stark model, here we also take into account the quadratic correction to the potential $\propto \alpha$. As shown in Appendix~\ref{sec:Appendix}, for $|\alpha|\ll |F|$ the system Eq.~\eqref{eq:stark} has an eigenvalue
\begin{equation}\label{eq:eigstark}
\varepsilon=-2\alpha\frac{\tau^2}{F^2}\approx -0.98 \left(\frac{t_2}{t_1}\right)^2={\rm const}(n_0)\:,
\end{equation}
corresponding to the state of the Wannier-Stark ladder localized at the site $n=n_0$. Importantly, the eigenenergy Eq.~\eqref{eq:eigstark}  does not depend on the column height $n_0$ and thus contributes to formation of a set of states differing by the value of $n_0$. These are exactly the states in Fig.~\ref{fig:eig}(b) shown by the diamond, square, circle and down-pointing triangle. Their energies are given by $|E|=t_1-t_2^2/t_1$, in agreement with Eq.~\eqref{eq:eigstark}.  This expression is plotted by the parabola with big red dots in Fig.~\ref{fig:levels} and perfectly agrees with the result of exact numerical calculation. 

Similar analysis can be made for the other DOS  peaks. We show several values of 
$k^{(\nu)}$, namely $k^{(\nu)}=\pi/4,3\pi/8,2\pi/5,3\pi/7,4\pi/9$,  and the corresponding energies $2t_1\cos k^{(\nu)}$  by the horizontal arrows in Fig.~\ref{fig:levels}. 
This Wannier-Stark-like model describes both localization and degeneracy of the spectrum in the limit when $t_2\ll t_1$.  For larger value of $t_2$ the coupling constant $\tau$ increases and the localization becomes weaker, but the model remains qualitatively correct. At the same time, due to the coupling between center-of-mass and relative degrees of freedom, the two particles become bound to each other, as can be seen from comparison of  the states shown by diamonds and squares in Fig.~\ref{fig:eig}(b) and Fig.~\ref{fig:eig}(d). At small values of $t_2$ the dependence of the energies on $t^{(2)}$ is parabolic and can be calculated by the perturbation series, $E(t_2)=E(t_2=0)-\nu t_2^2/t_1$. The corresponding expressions are plotted in Fig.~\ref{fig:levels} as dotted parabolas,  with the coefficient $\nu =2,1,2/3,1/2,1/3,1/4$ corresponding to the parabolas from top to bottom.

\section{Summary}
To summarize, we have considered a discrete two-particle problem in a simplest one-dimensional tight-binding model and demonstrated that when the particles are not identical, complex correlated states can result from their interaction, that can be linked to a non-Hermitian topological physics. It remains  to be understood what happens in the many-body  case and in  more complicated lattices. 
%\cite{Rhim2021}

%Topological 
%Reviews topo
%\cite{Rhim2021}
%
%
%
%
%Two particles
%\cite{Zhong2020}
%\cite{alex2020quantum,sheremet2021waveguide}
%\cite{Vidmar2022}
%\cite{Gorlach2017,Olekhno2020}
%\cite{Corrielli2013}
%\cite{Winkler2006}
%
%Reviews topo nonherm 
%\cite{Okuma2022}
%\cite{Bergholtz2021}
%\cite{Okuma2020}
%\cite{Kawabata2019}
%

\appendix
\section{Perturbation theory}\label{sec:Appendix}
Here we analyze  in more detail the Wannier-Stark-like model  Eqs.~\eqref{eq:stark} in the limit where $|\alpha|\ll 1$. We first start from the simpler case where also $|\tau|\ll 1$. The energy of the eigenstate, that is localized at the site $n=n_0$, can be then found by a simple second-order perturbation theory in $\tau$,
\begin{equation}\label{eq:naive}
\eps\approx \frac{\tau^2}{F-\alpha}+\frac{\tau^2}{F+\alpha}\approx \frac{2\alpha \tau^2}{F^2}\:.
\end{equation}
Importantly, since $\alpha/F^2$ does not depend on $n_0$, the states for different values of $n_0$ will be degenerate over the index $n_0$ describing the point where the first particle is localized. 

The interesting finding is that the answer Eq.~\eqref{eq:naive} remains valid even if $\tau\sim 1$, provided that still $|\alpha|\ll 1$. To this end we account for the term $\alpha (n-n_0)^2$   in Eqs.~\eqref{eq:stark} by a first-order perturbation theory in $\alpha$. For $\alpha=0$ Eqs.~\eqref{eq:stark} have eigenstates 
\begin{equation}\label{eq:J0}
\psi_n=J_n(-2\tau/F)\:.
\end{equation}
The first order perturbation theory correction is given by
\begin{equation}\label{eq:sum1}
\alpha\sum\limits_{n=-\infty}^\infty |J_n(-2\tau/F)|^2 n^2\:.
\end{equation}
We will now show that this sum is exactly equal to  Eq.~\eqref{eq:naive} for an arbitrary value of $\tau/F$. 
In order to prove this we explicitly use of the fact that Eq.~\eqref{eq:J0} is an eigenstate of Eqs.~\eqref{eq:stark} for $\alpha=0$. This means that 
\begin{multline}\label{eq:sum1}
\alpha n^2J_n^2(-2\tau/F)=\frac{\alpha\tau^2}{F^2} [J_{n-1}(-2\tau/F)+J_{n+1}(-2\tau/F)]^2\\=
\frac{\tau^2}{F^2} [J_{n-1}^2(-2\tau/F)+J_{n+1}^2(-2\tau/F)\\+2J_{n-1}(-2\tau/F)J_{n+1}(-2\tau/F)]\:.
\end{multline}
The summation over $n$ can be now performed analytically. First two terms in the square brackets in the right-hand side yield unity because of the normalization condition 
$\sum_nJ^2_n(-2\tau/F)=1$. The last term is zero because 
$J_{n+1}(-2\tau/F)$ and $J_{n-1}(-2\tau/F)$ are two different eigenstates of the Wannier-Stark problem Eq.~\eqref{eq:stark} with $\alpha=0$, hence  they are orthogonal to each other.  The result is Eq.~\eqref{eq:naive}.

%%%%%%%%%%%%%%%%%%%%%%%%%%%%%%%%%%%%%%%%%%%%%%%%%%%%%
%\begin{figure}[t]
%\centering\includegraphics[width=0.35\textwidth]{fig-stark.pdf}
%\caption{Schematic models to describe localization of interacting particles by formation of Wannier-Stark-like ladder. 
%(a) Coupled ``columns'' in the effective 2D model. (b) Interaction-induced potential for the states in such columns.}
%\label{fig:stark}
%\end{figure}
%%%%%%%%%%%%%%%%%%%%%%%%%%%%%%%%%%%%%%%%%%%%%%%%%%%%%

\begin{acknowledgements}
I am grateful to A.V. Poshakinskiy and  I.V. Rozhansky for useful discussions.
I thank  the Weizmann Institute of Science for  hosting me.
\end{acknowledgements}

%%%%%%%%%%%%%%
%\nocite{apsrev41Control}
%\bibliographystyle{apsrev4}
%\bibliography{topobound}

\end{document}